\documentclass[twocolumn]{aastex62}

\submitjournal{PASP}
\shortauthors{Wang Rui et al.}

\begin{document}

\title{Analysis of Stellar Spectra from LAMOST DR5 with Generative Spectrum Networks}

\correspondingauthor{Luo A-Li}
\email{lal@bao.ac.cn}

\author{Wang Rui}
\affil{National Astronomical Observatories, Chinese Academy of Sciences,Beijing 100012, China}
\affil{University of Chinese Academy of Sciences, Beijing 100049, China}
\affil{Key Laboratory of Optical Astronomy, National Astronomical Observatories, Chinese Academy of Sciences,Beijing 100012, China}

\author[0000-0001-7865-2648]{Luo A-li}
\affil{National Astronomical Observatories, Chinese Academy of Sciences,Beijing 100012, China}
\affil{Key Laboratory of Optical Astronomy, National Astronomical Observatories, Chinese Academy of Sciences,Beijing 100012, China}

\author{Zhang Shuo}
\affil{National Astronomical Observatories, Chinese Academy of Sciences,Beijing 100012, China}
\affil{University of Chinese Academy of Sciences, Beijing 100049, China}
\affil{Key Laboratory of Optical Astronomy, National Astronomical Observatories, Chinese Academy of Sciences,Beijing 100012, China}

\author{Hou Wen}
\affil{National Astronomical Observatories, Chinese Academy of Sciences,Beijing 100012, China}
\affil{Key Laboratory of Optical Astronomy, National Astronomical Observatories, Chinese Academy of Sciences,Beijing 100012, China}

\author{Du Bing}
\affil{National Astronomical Observatories, Chinese Academy of Sciences,Beijing 100012, China}
\affil{Key Laboratory of Optical Astronomy, National Astronomical Observatories, Chinese Academy of Sciences,Beijing 100012, China}

\author{Song Yihan}
\affil{National Astronomical Observatories, Chinese Academy of Sciences,Beijing 100012, China}
\affil{Key Laboratory of Optical Astronomy, National Astronomical Observatories, Chinese Academy of Sciences,Beijing 100012, China}

\author{Wu Kefei}
\affil{National Astronomical Observatories, Chinese Academy of Sciences,Beijing 100012, China}
\affil{Key Laboratory of Optical Astronomy, National Astronomical Observatories, Chinese Academy of Sciences,Beijing 100012, China}

\author{Chen Jianjun}
\affil{National Astronomical Observatories, Chinese Academy of Sciences,Beijing 100012, China}
\affil{Key Laboratory of Optical Astronomy, National Astronomical Observatories, Chinese Academy of Sciences,Beijing 100012, China}

\author{Zuo Fang}
\affil{National Astronomical Observatories, Chinese Academy of Sciences,Beijing 100012, China}
\affil{Key Laboratory of Optical Astronomy, National Astronomical Observatories, Chinese Academy of Sciences,Beijing 100012, China}

\author{Qin Li}
\affil{National Astronomical Observatories, Chinese Academy of Sciences,Beijing 100012, China}
\affil{University of Chinese Academy of Sciences, Beijing 100049, China}
\affil{Key Laboratory of Optical Astronomy, National Astronomical Observatories, Chinese Academy of Sciences,Beijing 100012, China}

\author{Chen Xianglei}
\affil{National Astronomical Observatories, Chinese Academy of Sciences,Beijing 100012, China}
\affil{University of Chinese Academy of Sciences, Beijing 100049, China}
\affil{Key Laboratory of Optical Astronomy, National Astronomical Observatories, Chinese Academy of Sciences,Beijing 100012, China}

\author{Lu Yan}
\affil{National Astronomical Observatories, Chinese Academy of Sciences,Beijing 100012, China}
\affil{University of Chinese Academy of Sciences, Beijing 100049, China}
\affil{Key Laboratory of Optical Astronomy, National Astronomical Observatories, Chinese Academy of Sciences,Beijing 100012, China}

\begin{abstract}
In this study, the fundamental stellar atmospheric parameters ($\textit{T}_{eff}$, log $g$, [Fe/H] and [$\alpha$/Fe]) were derived for low-resolution spectroscopy from LAMOST DR5 with Generative Spectrum Networks (GSN). This follows the same scheme as a normal artificial neural network with stellar parameters as the input and spectra as the output. The GSN model was effective in producing synthetic spectra after training on the PHOENIX theoretical spectra. In combination with Bayes framework, the application for analysis of LAMOST observed spectra exhibited improved efficiency on the distributed computing platform, Spark. In addition, the results were examined and validated by a comparison with reference parameters from high-resolution surveys and asteroseismic results. Our results show good consistency with the results from other survey and catalogs. Our proposed method is reliable with a precision of 80 K for $\textit{T}_{eff}$, 0.14 dex for log $g$, 0.07 dex for [Fe/H] and 0.168 dex for [$\alpha$/Fe], for spectra with a signal-to-noise in $g$ bands (SNR$_{g}$) higher than 50. The parameters estimated as a part of this work are available at \url{http://paperdata.china-vo.org/GSN_parameters/GSN_parameters.csv}
\end{abstract}


\section{Introduction}           
\label{sect:intro}

Most sky surveys result in extensive databases of stellar spectra for dissecting and understanding the Milky Way. The fundamental information derived from such spectra includes the effective temperature ($\textit{T}_{eff}$),  logarithm of surface gravity (log $g$), abundance of metal elements with respect to hydrogen ([Fe/H]), and the abundance of alpha elements with respect to iron ([$\alpha$/Fe]), are valuable for Galactic archaeology and stellar evolution history. Many projects have been carried out to detect specific objects at high/low resolution covering a range of wavelengths (e.g., the Large Sky Area Multi-Object Fiber Spectroscopic Telescope (LAMOST) Experiment for Galactic Understanding and Exploration ~\citep[LEGUE;][]{2015RAA....15.1095L,    2012RAA....12.1197C,    2012RAA....12..735D,  2012RAA....12..723Z,  2015RAA....15.1089L}, the Sloan Extension for Galactic Understanding and Exploration ~\citep[SEGUE;][]{2009AJ....137.4377Y}, the Apache Point Observatory Galactic Evolution Experiment~\citep[APOGEE;][]{2008AJ....136.2070A, 2017AJ....154...94M}, the RAdial Velocity Experiment ~\citep[RAVE;][]{2006AJ....132.1645S}, Gaia-ESO Public Spectroscopic Survey~\citep[GES;][]{2012Msngr.147...25G}, the GALAH Survey~\citep{2015MNRAS.449.2604D} and Gaia-RVS~\citep{2004MNRAS.354.1223K}).

Automatic and accurate estimation of stellar parameters from large spectroscopic survey databases requires a wide array of different techniques and methods. Almost each spectroscopy survey results in the establishment of its own stellar parameter pipeline, such as: the official LAMOST Stellar Parameter Pipeline~\citep[LASP;][]{2015RAA....15.1095L,  2011RAA....11..924W} based on UlySS package~~\citep{2009A&A...501.1269K}, LAMOST Stellar Parameter Pipeline at PKU~\citep[LSP3;][]{2015MNRAS.448..822X,2017MNRAS.464.3657X, 2016RAA....16...45R} which combines spectral fitting with an empirical spectral library and the application of Kernel-based Principal Component Analysis (KPCA) on different training sets to obtain parameters of interest, SEGUE Stellar Parameter Pipeline~\citep[SSPP;][]{2008AJ....136.2022L} using multiple techniques (e.g., spectral fitting with synthesis spectra grids, making use of comparisons to theoretical $ugr$ colors and line parameters from sythetic spectra\citep[WBG;][]{1999AJ....117.2308W} and neural network approaches), the APOGEE Stellar Parameter and Chemical Abundances Pipeline~\citep[ASPCAP;][]{2016AJ....151..144G} parametrizing near-infrared spectra by minimizing $\chi^{2}$ between observed and theoretical spectra, and the RAVE pipeline ~\citep{2006AJ....132.1645S} which exploits a best-matched template to measure radial velocities and atmospheric parameters. 

In these listed pipelines, the methods that most commonly used include spectral fitting based on grids of spectral libraries which can classify high/median-resolution observations based on empirical measurements, for example, ELODIE ~\citep{2007astro.ph..3658P}, STELIB~\citep{2003A&A...402..433L} and MILES ~\citep{2011A&A...532A..95F}, and theoretical synthetic spectra derived from atmospheric models such as: instance, ATLAS9 ~\citep{1993sssp.book.....K}, Kurucz2003~\citep{2004astro.ph..5087C}, MARCS ~\citep{2008A&A...486..951G} and PHOENIX~\citep{1995ApJ...445..433A, 2013A&A...553A...6H}. However, neither of these spectral libraries offer a perfect solution to the problem of deriving stellar atmospheric parameters. The empirical spectral libraries are limited in wavelength and parameter space coverage, for example, the latest version v3.2 of the ELODIE library lacks a template of K giant stars, and most libraries are incomplete or lack chemical abundance data. The synthetic spectra would introduce large systematic uncertainties, mainly because of the impertinent stellar atmospheric models, such as the assumptions of one-dimensional models and the fact that local thermodynamic equilibrium for hot and cool stars is not completely consistent with actual conditions. Also, it is computationally expensive to generate synthetic spectra starting from physical assumptions. However, synthetic libraries have many advantages: the range of stellar parameters, elemental abundances, and both the wavelength coverage and the spectral resolution can be adjusted as needed. In this work, we employed the least version of Phoenix~\citep{2013A&A...553A...6H}, and the details are provided in Section~\ref{sect:data}. 

The effective utilization of synthetic spectra information is a critical process in the analysis of observed spectra. The increase in the number of spectra with time, both synthetic and observed, provides an opportunity to apply various learning methods to derive stellar parameters, independent of a large number of synthetic spectra.~\citet{1997MNRAS.292..157B, 2000A&A...357..197B} performed the incipient study on the application of Artificial Neural Network(ANN) for spectral classification and parameter estimation. Subsequently, numerous related techniques have been explored such as the Non-linear regression model~\citep{2007A&A...467.1373R} and Support Vector Regression+LASSO~\citep{2014ApJ...790..105L} for estimation of parameters from SDSS/SEGUE spectra, KPCA~\citep{2017MNRAS.464.3657X} and Cannon~\citep{2015ApJ...808...16N, 2017ApJ...841...40H} adopted an data-driven approach for the analysis of LAMOST stellar spectra, StarNet~\citep{2018MNRAS.475.2978F} and AstroNN~\citep{2018arXiv180804428L} applied  Convolutional Neural Networks technique to an APOGEE data-set, and MATISSE~\citep{2006MNRAS.370..141R}, ANN~\citep{2010PASP..122..608M} and Aeneas~\citep{2012MNRAS.426.2463L} employed machine learning method for Gaia RVS low S/N stellar spectra parametrization. All the aforementioned algorithms have their own unique strengths for constraint mapping from spectral flux to parameters (\textit{T}$_{eff}$, log \textit{g} and metal abundance). However, many of them take spectrum fluxes as inputs with the parameters and stellar labels as outputs of the algorithms. As such, they cannot directly determine the errors of the outputs because these methods cannot produce the post-distribution of the parameters for a given spectrum. The error information they generate is most relevant to the internal errors of each method or the dispersion of the difference between their results and that of external counterparts. Simple linear interpolation could not precisely characterize the complex relationship between flux and stellar parameters until~\citet{2016ApJ...826L..25R, 2016ApJ...826...83T, 2018arXiv180401530T} put forward a polynomial spectral model and the \emph{Payne} method, that are both based on the training of a mathematical model using ab initio spectral model grids. These approaches benefit from artificial neural networks because they are effective at fitting complex non-linear relations.

In this report, we designed a new structure of artificial neural networks, Generative Spectrum Networks(GSN), a similar neural network proposed by~\citet{2016A&A...594A..68D}, which follows the same scheme as a typical ANN, except that the inputs and outputs are inverted. This approach was proposed and applied to simulations of prospective Gaia RVS spectra based on the Kurucz model. However, real observed spectra were not tested. It should be noted that there is a sign discrepancy between the synthetic and observed spectra for errors from extinction, redden, seeing, contamination of stray light, instruments and post data processing. We improve the generative artificial neural network training on Phoenix spectra for estimation of the parameters of LAMOST observations. In combination with a Bayesian framework and Monte Carlo(MC) method, the networks can be used to derive not only stellar atmospheric parameters, but also their posterior distribution. The computing cost is always an insurmountable obstacle during the application of the MC method for a large number of data-sets. However, the distributed computing platform SPARK improves the viability of employing MC sampling methods based on Bayes theory. To the best of our knowledge, we are the first group to utilize SPARK estimating stellar parameters in this way. Moreover, our method adds an abundance of alpha elements ([$\alpha$/Fe]) with respect to the existing catalog provided by LASP.
 
This report is organized as follows. The Phoenix datasets for training and testing the GSN model, and the LAMOST DR5 observations will be described in Sect.\ref{sect:data}. The methods for the determination of the stellar atmospheric parameters will be presented in Sect. \ref{sect:methods and application}. Validation results will be highlighted in Sect. \ref{sect:validation}. Finally, we will discuss the challenges associated with this research in Sect. \ref{sect:discussion} followed by a summary in Sect. \ref{sect:summary}.  

\section{Data}
\label{sect:data}

The spectra employed in this report consist of two parts:  synthetic spectra calculated from PHOENIX mode~\citep{2013A&A...553A...6H} and LAMOST spectra from the internal fifth data release(LAMOST DR5; Luo et al. in preparation). The synthetic spectra with reference parameters(\textit{T}$_{eff}$, log \textit{g}, [Fe/H] and [$\alpha$/Fe]) are used for training and testing the GSN model. Then, the stellar parameters of the LAMOST spectra were estimated using the achieved GSN model.  

\subsection{Synthetic Spectra}

The synthetic stellar spectral library is a very important tool for analyzing observed spectra and stellar population synthesis. The latest extensive library of PHOENIX stellar atmospherics and high-resolution synthetic spectra are published in~\citet{2013A&A...553A...6H}. This new spectral library uses version 16 of PHOENIX, which includes a new equation of state, as well as an update-to-data atomic and molecular line list. The atmospheres were calculated based on a spherical mode with 64 layers, and both local thermodynamic equilibrium (LTE) and non-local thermodynamic equilibrium (N-LTE) have been adopted for an effective temperature \textit{T}$_{eff} \geq$ 4000 K. As a result, the synthetic spectra matching observation is significantly better than that of other synthetic libraries, especially for cool stars . Further details can be obtained in a published report~\citep{2013A&A...553A...6H}. The library we used contained 27,700 spectra, and their parameter space coverage and grid are shown in Table \ref{Tab:param grid}. The version we adopted archives with full sub-grids of the library, convoluted to the resolution $\Delta\lambda = 1$ \AA ~in the optical wavelength range from $\lambda = 3000$ \AA ~to $10000$ \AA ~and over-sampled by a factor of ten (the 1 \AA ~grid has a sampling rate of 0.1 \AA)~\citep{2013A&A...553A...6H}.
 
\begin{table}[!h]
\begin{center}
\centering
\caption{ Parameter Space of the Grid.}
\label{Tab:param grid}
\begin{tabular}{clcl}
\hline\noalign{\smallskip}
Variable  &  Range  &  Step size                    \\
\hline\noalign{\smallskip}
\textit{T}$_{eff}$ &   2300 - 7000  &  100                                \\
	                         &    7000 - 12000 &  200                               \\	                               
log \textit{g}  & 0.0 - +6.0    &    1.0                                        \\
\ [Fe/H]  &  -4.0 - 2.0  &  1.0                                                       \\
	          &  -2.0 - +1.0  &  0.5                                                     \\
\ [$\alpha$/Fe]  &  -2.0 - +1.2  &  0.2                                           \\
\noalign{\smallskip}\hline
\end{tabular}
\end{center}
\end{table}

Before spiting the spectral library into training and testing sets, some adjustments were made to facilitate matching of low-resolution spectra. To fully exploit the spectral information hidden in the LAMOST spectra, we consider using the largest wavelength range ($\lambda = 3800$ to $5700$ \AA ~for the blue arm, and $\lambda = 5900$ to $8800$ \AA ~for the red arm, and we removed the middle range ($\lambda = 5700$ to $5900$ \AA) in which the response efficiency of instruments is too low). According to the instrumental full-with-half-maximum of the LAMOST spectrometers (as shown in Figure \ref{Figure1}), we respectively convolute the blue and red arms of the Phoenix model spectra to the same resolution as the LAMOST observations. Then, we simply normalized the spectra by dividing them by their medians. For their corresponding parameters, the effective temperature was adjusted in a logarithmic scale. 

\subsection{Observed Spectra from LAMOST DR5}

\begin{figure}[tbp]
\centering
\includegraphics[width=0.5\textwidth, angle=0]{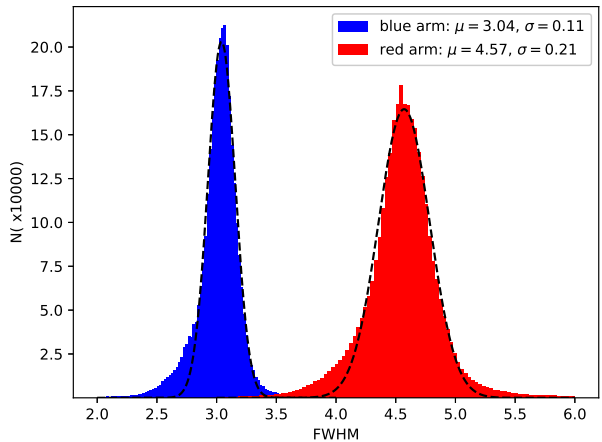}
\caption{The distribution of the instrumental full-width-half-maximum(FWHM) of the blue (blue) and red (red) arm for LAMOST. The mean and standard deviation of the FWHM of the blue arm are 3.04 \AA ~and 0.11 \AA ~respectively. For red the arm, the mean and standard deviation are 4.57 \AA ~and 0.21 \AA , respectively.}
\label{Figure1}
\end{figure}

The Large Sky Area Multi-Object Fiber Spectroscopic Telescope (LAMOST) general survey is a spectroscopic survey~\citep{2015RAA....15.1095L} which has already completed the first stage of the sky survey plan (from Oct. 24th, 2011 to Jun. 16th, 2017). It contains two main parts: the LAMOST ExtraGAlactic Survey (LEGAS) and the LAMOST Experiment for Galactic Understanding and Exploration (LEGUE) survey of the Milky Way stellar structure. The LAMOST can simultaneously collect
4000 fiber spectra in a wild field ($5^{\circ}$) with a resolving power R$ \approx $1800. Each of the final stellar spectra is a combination of three single-exposure spectra which consist of two sub-spectra obtained in the blue (wavelength range: 3800-5900\AA) and red (wavelength range: 5700-9000\AA) arms of the spectrometers. The instrumental full-width-half-maximum of the two arms are not exactly the same. As such, the spectra of each arm is treated separately (see Figure \ref{Figure1}). All released spectra are wavelength calibrated at a vacuum wavelength based on lamp spectra and represent the relative flux calibrated based on standard stars~\citep{2015RAA....15.1095L} or statistical response curve within an error of 10\% for each pixel~\citep{2016ApJS..227...27D}. The spectra are re-binned at a constant interval wavelength for a logarithmic scale.

In this work, we exploit artificial neural networks for the analysis of LAMOST DR5 stellar spectra. The fifth data release of LAMOST includes a total of 4,154 plates containing 9,017,844 spectra, of which 8,171,443 are classified as stars and the rest as galaxies, quasars, or unknown objects via the LAMOST 1D pipeline. The number of stellar spectra with a signal to noise ratio (SNR) of the g or i band higher than ten is 7,531,398. LAMOST DR5 also provide three stellar catalogues: late A and FGK-type stars with high quality spectra (5,344,058) with measured parameters(\textit{T}$_{eff}$, log \textit{g}, [Fe/H], radial velocities (RV)) derived by LASP ~\citep{2011RAA....11..924W, 2015RAA....15.1095L}, early A-type stars (365,101 entries) and M-type stars (436,782 entries) without stellar parameters. 

\section{Method and Application}
\label{sect:methods and application}

\subsection{Generative Spectrum Networks}

Similar to~\citet{2000A&A...357..197B} and~\citet{2010PASP..122..608M}, many researcher design traditional artificial neural networks structures with the fluxes of the spectra as the input and the stellar atmospheric parameters as the output, for analysis of spectra. ~\citet{2016A&A...594A..68D} suggested the innovative idea of inverting the inputs and outputs of the ANNs, with the parameters as inputs and the flux values as outputs. The \emph{Payne}~\citep{2018arXiv180401530T} also adopted an analogous, fully connected neural network to produce reference spectra instead of an interpolate operator. These kinds of neural networks perform a precise fitting of the non-linear relationship between fluxes and stellar parameters. 

In this report, we designed a new structure of artificial neural networks which consists of a fully-connected network with an input layer, three hidden layers, and an output layer, to generate spectra by training PHOENIX model spectra. The name of the networks was changed from generative artificial neural networks (GANN) to Generative spectrum networks (GSN) for differentiating with generative adversarial networks (GAN). The connection between all neurons in two layers is constructed by non-linear combinations, at each layer $l+1$, and the output as a function of the previous layer $l$ is given by: 
 \\
 \\
$ y_j^{l+1}=g( \sum\limits_{i=0}^n (w_{ij}^l y_{ij}^l+b_{j}^l) $, 
\\
\\
where $g$ is an activation function, $w_{ij}$ is the weight representing the connection of the node $i$ of layer $l$ and node $j$ of layer $l+1$, and $b_{ij}^l$ is the bias of the node $i$ of layer $l$. There are many choices options for the activation function, such as ReLU, Sigmoid, Logistic, Linear etc. All these activation functions can change the linear relationship between the layers except for the Linear activation function. We chose ReLU for the middle hidden layers and the linear activation function for the output layer. The Training model for optimizing the weights and biases is based on approaches such as stochastic gradient descent (SGD), mining error functions $E$:
\\
\\
 $E=\displaystyle\frac{1}{n}\sum\limits_{i=0}^n (y_{i}^l- \hat{y_{i}^{l}})^2$   
\\
\\
where $\hat{y_{i}^l}$ is the prediction outcome of ANN model and $y_{i}^l$ is the actual values of the fluxes of the training set.

Training the neural network described above is based on a back propagation (BP) algorithm in which the parameter weights and biases are initialized with random values. However, vanishing and exploding gradient will occur when the number of hidden layers is too large. Here, we introduce an auto-encoder to initialize the weights and biases and then fine-tune to address the gradient problem described above~\citep{HintonSalakhutdinov2006b}. Also, we employ early-stopping to prevent over-fitting from occurring when the number of training samples is much less than the complexity of the network structure. We randomly split the synthetic spectra into the training set, validation set and test set in the proportion 5:1:4, and train different structures of GSN to obtain the best one among them. Finally, a five-layer network, consisting of the input layer of 4 stellar parameters, three hidden layers with 40, 400, 1000 nodes, and the final output layer of 3641 spectral fluxes corresponding to the wavelength points of our re-sampled LAMOST spectra, is chosen because of its well-documented high performance.

\subsection{Fully Bayesian}
 
Generative Spectrum Networks can produce model spectra when a team of parameters is given. Using chi-square distance as a proxy to match the spectrum to be measured with model grids is common and most methods use this approach. However, the uncertainty estimation would be difficult for template matching. Combined with Bayes rule, Monte-Carlo sampling is an effective way to obtain the posterior distribution over the parameters given the observed spectrum, $P(Param|Spec)$, as represented by the formula (6) in~\cite{2016A&A...594A..68D} 
\\
\\
$P(Param|Spec)=\displaystyle\frac{P(Spec|Param)P(Param)}{P(Spec)}$,
\\
\\
where $P(Spec|Param)$ is the probability of the result parameters $Param$ when a spectrum $Spec$ to be estimated is given, $P(Param)$ is the prior distribution of the parameters, $P(Spec)$ is a normalization factor. Here, we assume that $P(Param)$ is a uniformly distributed and the noise distribution function of the observed spectrum is a Gaussian-like function although the main noise that comes from the CCD read noise satisfies a Poison distribution. Therefore, the $P(Spec|Param)$ can be expressed as:
\\
\\
$P(Spec|Param)=\\
\displaystyle exp\{-\frac{1}{2}{\sum\limits_{i=1}^{N}(\frac{F_{obs,i}-Poly_{n,i} \times F_{GSN,i}}{\sigma_{i}})^{2}}\}$,
\\
\\
where $F_{obs,i}$ is the flux of the pixel $i$ of the observed spectrum $Spec$, $Poly_{n,i} $ is a $n$ order polynomial correction adjustment item for the uncertainty of flux calibration and redden effect, $F_{GSN,i}$ is the flux of the pixel $i$ of the generative spectrum of the GSN model when the input parameters is $Param$ and $\sigma_{i}$ is the error of the observed flux.

\subsection{Application to LAMOST DR5}

We estimated stellar parameters (\textit{T}$_{eff}$, log \textit{g}, [Fe/H] and [$\alpha$/Fe]) of LAMOST DR5 $\sim $5,300,000 spectra with SNR$_{g}>30$ using GSN and Bayesian methods as described above. All stellar spectra are initially shifted to their rest-frame wavelength using the radial velocity provided by LAMOST 1D pipeline and re-sampled to a fixed wavelength range. To obtain the posterirori distribution of the parameters containing the 'real' values, 5000 groups parameters were sampled from the uniform distribution $ U[\mu_{i}-3\sigma_{i}, \mu_{i}+3\sigma_{i}]$, where $i$ denote parameter(\textit{T}$_{eff}$, log \textit{g}, [Fe/H] or $[\alpha/Fe]$) provided by LASP, $\mu_{i}$ and $\sigma_{i}$ are parameters and their corresponding errors. LASP do not supply the parameter $[\alpha/Fe]$. In this case, $[\alpha/Fe]$ is set as~\cite{2013AJ....146..132L}: $\alpha$-enhancement ratio relative to Fe, is $-0.4\times x$\ for $x= $ [Fe/H], and the standard deviation is set to 0.3. Previous researchers have determined that the log $g$ supplied by LASP is not accurate for K giants~\citep{2014ApJ...790..110L, 2015RAA....15.1125C, 2016AJ....152....6W}. As such, we calibrated the log $g$ supplied by LASP based on the calibration relation given by equation (12) of~\citet{2016AJ....152....6W}:
\\
\\
$log\ g(adopted)=2.188\times log\,g({lasp})-0.2882\times log\,g({lasp}) \times \displaystyle\frac{T_{eff}({lasp})}{1000} +1.283\times \displaystyle\frac{T_{eff}({lasp})}{1000} -5.716$
\\
\\
for stars with $3800 K\leq Teff\leq 4500 K$,  $1.3\leq log g \leq 2.2$, or $3800 K\leq Teff\leq 5200 K$, $2.2\leq log g \leq 3.5 $.

The excessive computing time associated with Monte Carlo progress combining Bayes rule for large numbers of data can be a significant hindrance. In this case, we make use of a very efficient distributed-computing platform, SPARK, which considerably reduces computing time. Each spectrum was sampled  5000 times which results in more than 5.3 million spectra from LAMOST DR5. The total time required is only 40.5 hours using a SPARK computer cluster consisting of 16 computers after formatting our data as a Hadoop Distributed File System (HDFS) data storage format.

\section{Results and Validation}
To ensure the reliability and accuracy of the stellar parameters obtained with GSN, we employed the parameter catalogs of some sub-sample catalogs of LAMOST DR5 common stars, with external precise stellar parameters derived from high-resolution observations, or by other methods used for comparing and validation. To obtain reliable results, we only selected spectra with SNR$g>$30 for comparison purposes. 
\label{sect:validation}

\subsection{Comparison with APOGEE Stars}

The Apache Point Observatory for Galactic Evolution Experiment~\citep[APOGEE;][]{2015AJ....150..148H, 2017AJ....154...94M, 2018AJ....156..125H} is a median-high resolution (R$\sim$22 500) spectroscopic survey in the near-infrared spectral range ($\lambda = 15700$ to $17500$ \AA). From SDSS DR14, APOGEE has already collected $\sim$277,000 spectra which are predominantly giant stars and are parameterized based on the APOGEE Stellar Parameters and Chemical Abundances Pipeline~\citep[ASPCAP;][]{2016AJ....151..144G}, based on a comparison between the observed and theoretical spectra using chi-square minimization. The complete SDSS DR14 APOGEE sample has 34783 stars in common with the LAMOST DR5 sample which correspond to 146,697 spectra. We neglected the dwarf stars(log $\ge3.5$) for which APOGEE did not provide reliable parameters, in addition to the stars flagged by APOGEE as having possible problems in the spectrum or in the ASPCAP progress. This finally left us with a total of  21,642 spectra after the exclusion of repeat observations and the LAMOST spectra with SNR$g\le$ 30.

\begin{figure*}
\includegraphics[width=\textwidth, angle=0]{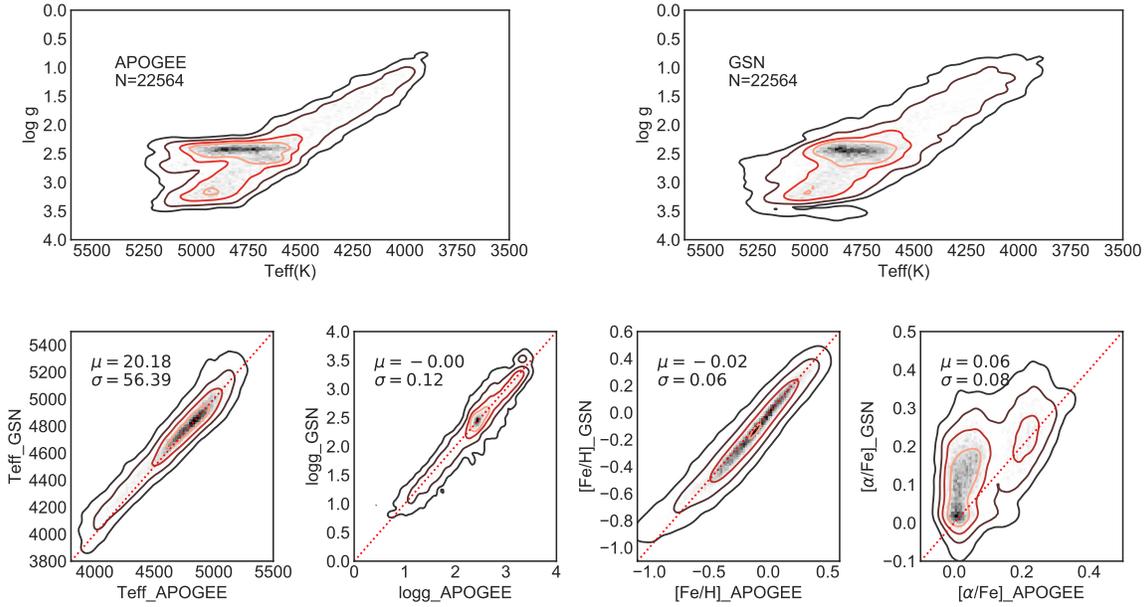}
\caption{The density distribution of APOGEE parameters (top left panel) and GSN results (top right panel) for LAMOST 23,315 stars with SNR$_{g}$ greater than 30. Also, the comparison between GSN stellar parameters and the APOGEE parameters are showed in the four bottom panels. The red dash lines in the bottom panels are one-to-one lines.}
\label{Figure2}
\end{figure*}

A comparison of results for the common stars observed by both LAMOST and APOGEE is shown in Fig.~\ref{Figure2}. The top panels show the distribution of the stars in the ($T_{eff}$, log$g$) plane and the bottom panels show the detail contradistinction for each parameter. The results indicate good agreements between the two parameter sets with small offsets and scatter. The offsets of degeneracies are 20.18 K, 0.00 dex, -0.02 dex, 0.06 dex and the dispersions are 56.39 K, 0.12 dex, 0.06 dex, 0.08 dex for $\textit{T}_{eff}$, log \textit{g}, [Fe/H], [$\alpha$/Fe], respectively. It should be noted that there is a small degree of systematic offsets. However, the small dispersions suggest that the estimations are precise. The temperature derived from the GSN exhibits an overestimation of the APOGEE results by 20.18 K. In addition, the estimated [Fe/H] by GSN shows an underestimation of 0.02 dex with respect to the [Fe/H] of APOGEE, while [$\alpha$/Fe] overestimates by 0.06 dex. However, it is worth noting that the stellar atmospheric parameters derived by ASPCAP are calibrated~~\citep{2015AJ....150..148H, 2018AJ....156..125H}. The raw ASPCAP temperature which is determined to be approximately 90K cooler than the photometric temperature is calibrated by a linear fit relation derived by inferred color-temperature. However, our temperature is still unexpectedly higher by approximately 20 K compared to that of APOGEE. The calibration relation for the surface gravity was based on a set of APOKASC catalog ~\citep{2014ApJS..215...19P}, therefore the APOGEE log \textit{g}s were scaled to the asteroseismic results. In this case, no systemic offsets were identified for the surface gravity between GSN and ASPCAP, which benefit from the relatively accurate initial value coverage for GSN estimation. For [$\alpha$/Fe], the APOGEE dataset exhibits a small offset of 0.06 dex, with a small scatter of 0.08 dex, which would be considered as a precise result for such low-resolution spectra. This result is achieved because the alpha-to-iron ratio of the APOGEE stars is measured based on the spectra with a resolution that is 11 times higher.  Moreover, the GSN measures abundance by computing the posterior probability from the information in the entire spectral bands (3800-8800\AA), which may lead to a bias of 0.1-0.2 dex~~\citep{2015AJ....150..187L}. We can shift the [$\alpha$/Fe] to the APOGEE's values to eliminate system bias.

\subsection{Comparison with PASTEL Catalogue}

The PASTEL catalogue~\citep{2010A&A...515A.111S,2016A&A...591A.118S} is a bibliographical compilation of stellar parameters ($\textit{T}_{eff}$, log \textit{g}, [Fe/H]) which were obtained from the analysis of high resolution and high signal-to-noise spectra, derived based on various methods.~\citet{2016A&A...591A.118S} updated the PASTEL catalogue which includes 64,082 determinations of either $\textit{T}_{eff}$ or ($\textit{T}_{eff}$, log \textit{g}, [Fe/H]) for 31,401 stars, of which 11,197 stars have a values for the three parameters ($\textit{T}_{eff}$,log \textit{g}, [Fe/H]) with a high-quality spectroscopic metallicity. These stars can therefore be used as reference to compare with our results. Our LAMOST sample has 1,000 common stars with the PASTEL catalogue, of which there are 579 with $\textit{T}_{eff}$, 372 with log \textit{g} and 565 with [Fe/H], after excluding the outliers and low signal-to-noise spectra using the same criteria as~\cite{2015RAA....15.2204G}: 
\\

$|T_{eff}(GSN) - T_{eff}(PASTEL)|\geq1000 $ K

$|log~g(GSN) - log~g(PASTEL)|\geq0.5 $ dex

$|[Fe/H](GSN) - [Fe/H](PASTEL)|\geq0.5 $ dex

SNR$_{g}(LAMOST)\leq30$
\\
\begin{figure*}[tbp]
\centering
\includegraphics[width=\textwidth, angle=0]{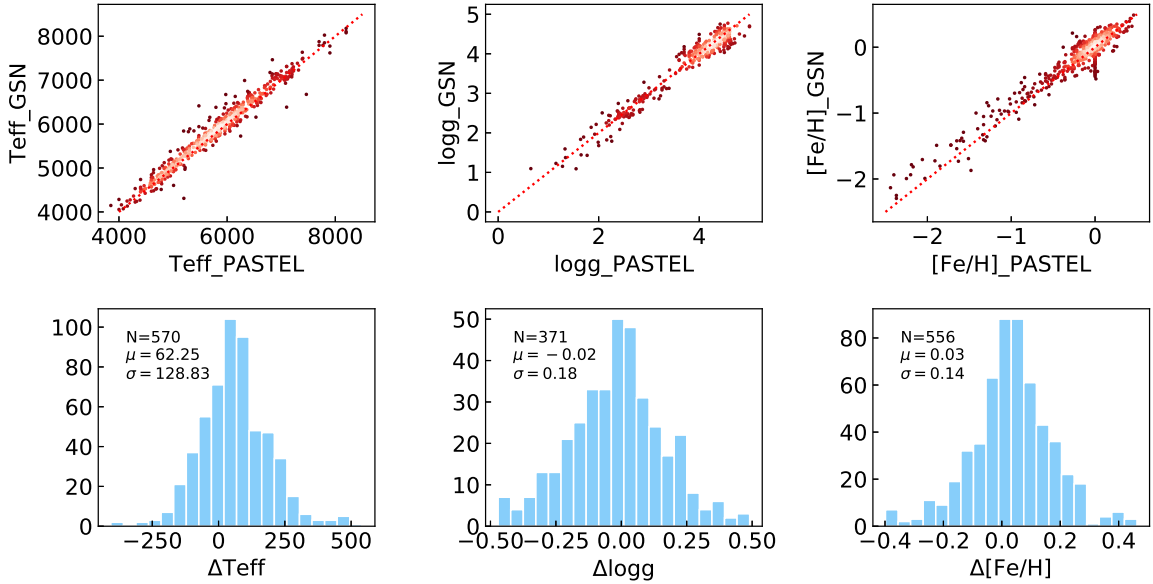}
\caption{The color-coded scatter of GSN stellar parameters compared with the PASTEL catalogue for LAMOST DR5 spectra with SNR$_g\geq30$ are showed in the three top panels. The red dash lines in the top panels are one-to-one lines. The distributions of the discrepancies for $\textit{T}_{eff}$, log $g$ and [Fe/H] are shown in the three bottom panels.}
\label{Figure3}
\end{figure*}

A comparison of results between the stellar parameters derived by GSN and that provided by the PASTEL catalog is shown in Fig~\ref{Figure3}. Most of the LAMOST-PASTEL common stars are dwarfs. The results indicate a good consistency between the parameters of GSN and that of the PASTEL catalogue: no significant offset was observed for log \textit{g}s or for [Fe/H]s except for $\textit{T}_{eff}$ with an offset of 62.25 K. The dispersion of their difference is given as 128.83 K, 0.18 dex and 0.14 dex respectively. ~\citet{2016A&A...591A.118S} noted that the fundamental Teff based on direct measurements of the angular diameters and the total flux of Earth is cooler than the temperature listed in the 2006 version of the PASTEL catalog by 51 K , with a median absolute deviation of 96 K. Our temperature overestimate of 62.25 K means that our temperature scale seems to be about 110 K higher than that of the fundamental temperature. 

\subsection{Comparison with Geneva-Copenhagen Survey Stars}

The Geneva-Copenhagen survey~\citep[GCS;][]{2004A&A...418..989N,2007A&A...475..519H} present 16,682 F and G dwarfs in the solar neighborhood with ages, metallicities and kinematic properties.~\citet{2011A&A...530A.138C} derived the effective temperature $\textit{T}_{eff}$ and gravity log \textit{g} of the GCS sources from photometry based on the infrared flux method (IRFM), and used the derived temperature to build a consistent metallicity scale. Subsequently, the metallicity was calibrated using high-resolution spectroscopy. Also, Casagrande derived a proxy for [$\alpha$/Fe] abundances from Str$\ddot{o}$mgren photometry. LAMOST have 553 stars in common with GCS, of which 366 have SNR$_{g}\geq30$. 

\begin{figure*}[tbp]
\centering
\includegraphics[width=\textwidth, angle=0]{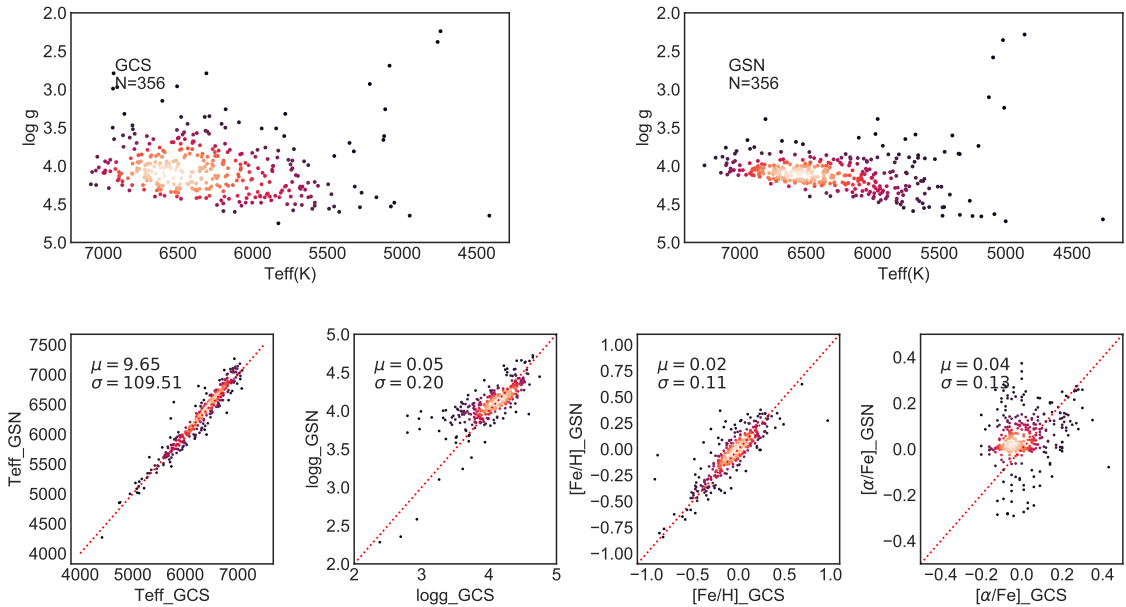}
\caption{The density distribution of GCS parameters (top left panel) and GSN results (top right panel) for 367 LAMOST-GCS common stars with SNR$_{g}$ of LAMOST spectra larger than 30. Also, the comparison between GSN stellar parameters and the GCS parameters derived by IRFM are showed in the four bottom panels. The red dash lines in the bottom panels are one-to-one line.}
\label{Figure4}
\end{figure*}

The comparison results between stellar parameters derived by GSN and IRFM ~\citep{2011A&A...530A.138C} are shown in Fig~\ref{Figure4}. From the comparison, we observed that our results match the parameters of GCS very well and the offsets are 9.65 K, 0.05 dex, 0.02 dex and 0.04 dex respectively for $\textit{T}_{eff}$, log \textit{g}, [Fe/H] and [$\alpha$/Fe], with small dispersions of 109.51 K, 0.20 dex, 0.11 dex and 0.13 dex, respectively. The encouraging contiguous results of GSN and GCS show great agreement for all the stellar atmospheric parameters. The offset of 9.65K of the surface temperature reminds us that our temperature scale is similar to IRFM's, which is known to be approximately 100K hotter than the spectroscopic temperature~\citep{2011A&A...530A.138C}. This is also observed above.

\subsection{Comparison with Asteroseismic Results}

\begin{figure*}[tbp]
\centering
\includegraphics[width=\textwidth, angle=0]{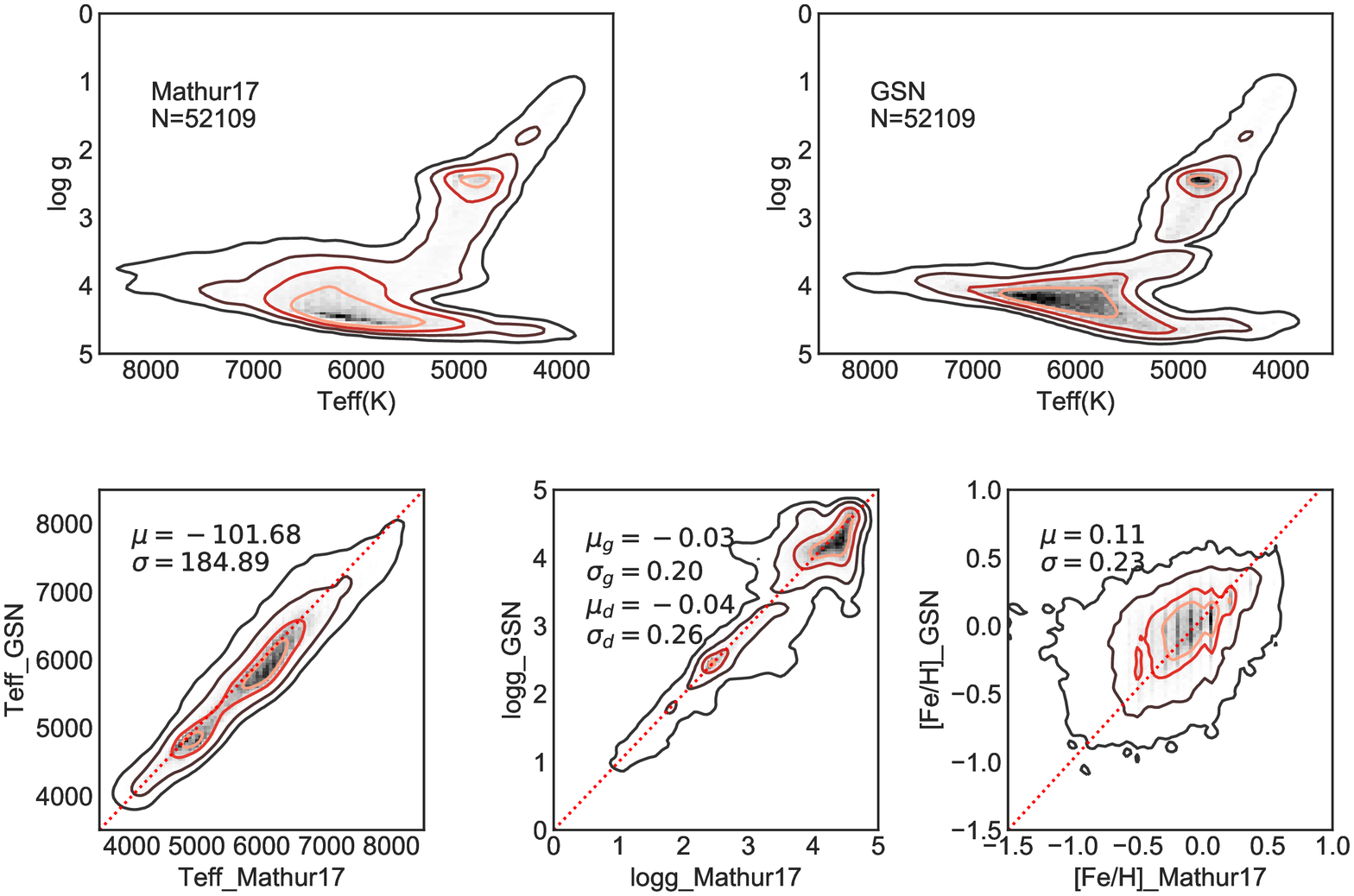}
\caption{The density distribution of Mathur17 parameters (top left panel) and GSN results (top right panel) for 52,109 LAMOST stars with SNR$_{g}$ larger than 30. Also, a comparison of GSN stellar parameters with the Mathur17 parameters are showed in the three bottom panels. The red dash lines in the bottom panels are one-to-one line.}
\label{Figure5}
\end{figure*}

The highly precise measurement of the asteroseismic gravity log \textit{g} has resulted in their frequent use as a reference for comparison and validation. The NASA Kepler mission has made available many host stars oscillations, which is the basis of asteroseismology.~\citet{2014ApJS..211....2H} (hereafter Huber14) reported a catalog, which consisted of a compilation of literature values for stellar parameters derived from photometry, spectroscopy, asteroseismology, or exoplanet transits information of 196,468 stars observed during the Kepler mission. Most parameters of Huber14 are more precise than the results provided by the Kepler Input Catalogue ~\citep[KIC;][]{2011AJ....142..112B}, however, the effective temperature and metallicities of Huber14 are still inaccurate.~\citet{2017ApJS..229...30M} revised the stellar properties of this catalog by improving some of the input parameters of spectroscopy and asteroseismology and correcting some of the parameters of misclassified stars. We employ the revised catalogue (~\citet{2017ApJS..229...30M}, hereafter Mathur17) for comparison of both giants (with $log g \le3.5$) and dwarfs (with $log g \ge3.5$).  The total number of Mathur17 is 197,096. We cross-match LAMOST DR5 and obtain 72,434 spectra, out of which 52,094 have SNR$_{g}\geq30$. The comparison results are shown in Fig~\ref{Figure5}. The offsets of temperature and [Fe/H] are still evident, which is not useful in the evaluation of the accuracy. Although Mathur17 improved many stellar properties, the KIC with a systematic overestimation of the temperature of 200 K and an underestimation of [Fe/H] of 0.1 dex, still takes up a large proportion as the input parameters. This may explain the offsets of the temperature and [Fe/H] shown in Fig~\ref{Figure5}. However, our surface gravity shows a good agreement with Mathur17's, both for the giants with a tiny offset of -0.03 dex and a scatter of 0.20 dex, and for the dwarfs with a small offset of -0.04 dex and a scatter of 0.26 dex. The scatters may arise primarily from the surface gravity input values of Mathur17 and the gravity uncertainty which was reported as associated uncertainties of 0.03 dex from seismology and 0.40 dex from the KIC ~\citep{2017ApJS..229...30M}.

Recently,~\citet{2018ApJS..236...42Y} (hereafter Yu2018) produced a homogeneous catalogue of asteroseismic stellar properties for 16,094 red giants. This provides an excellent opportunity to test the atmospheric parameters for large spectroscopic surveys using asteroseismic results. We also selected the Yu2018 sample as a reference to compare with our parameters, especially for validating log \textit{g}. There are 10,902 stars in common between Yu2018 and LAMOST, out which 8,647 have SNR$_{g}\ge30$. 

\begin{figure*}[tbp]
\centering
\includegraphics[width=\textwidth, angle=0]{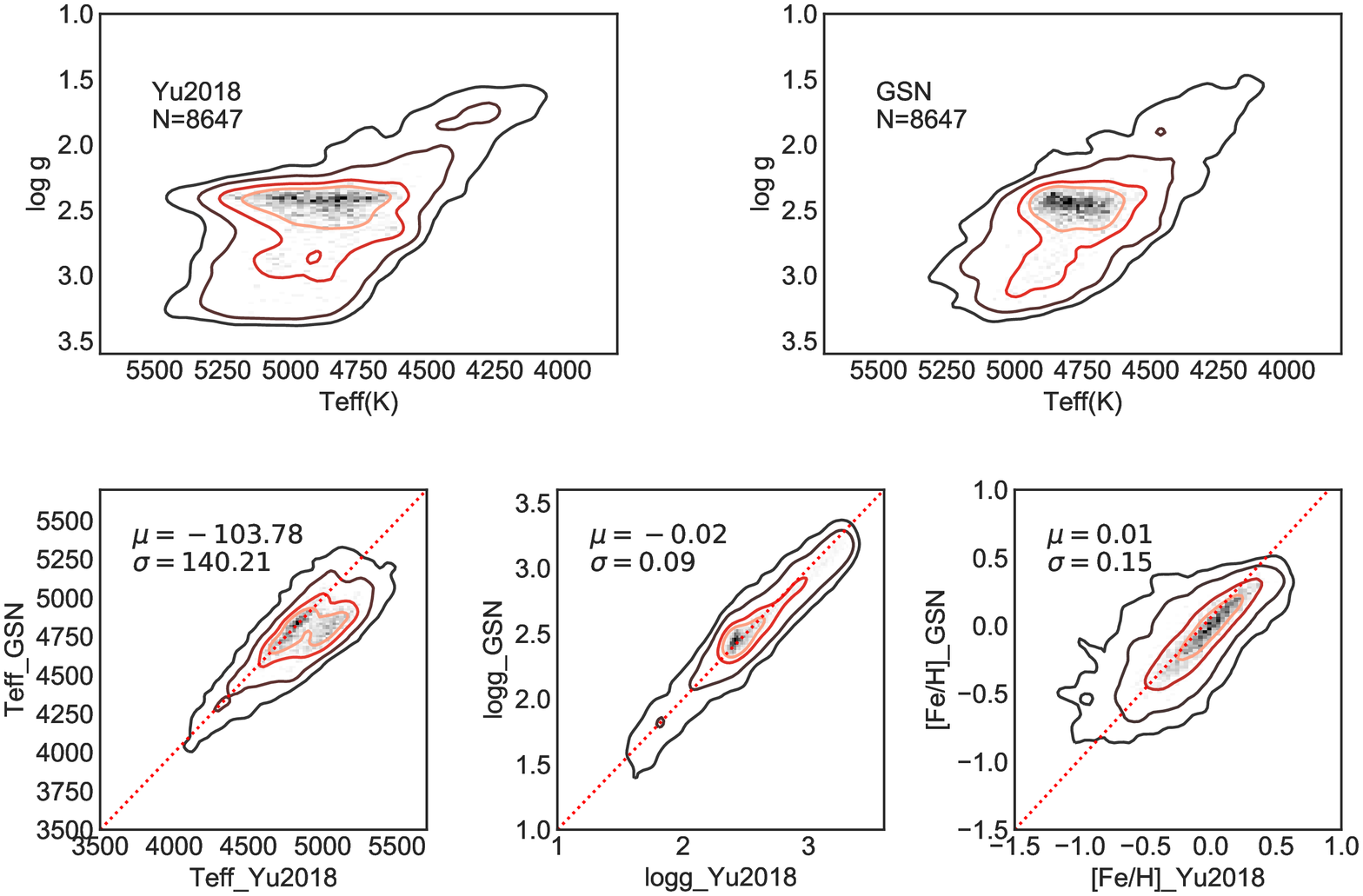}
\caption{The density distribution of Yu2018 parameters (top left panel) and GSN results (top right panel) for 8647 LAMOST stars with SNR$_{g}$ larger than 30. Also, a comparison of GSN stellar parameters with the Yu2018 parameters are showed in the three bottom panels. The red dash lines in the bottom panels are one-to-one line.}
\label{Figure6}
\end{figure*}

Fig~\ref{Figure6} shows a comparison of the results between the stellar parameters derived by GSN and Yu2018 using the same approach described above. From the comparison of the surface gravity, it is evident that our results are in good agreement with the asteroseismic results. In particular, for log \textit{g}, the offset of -0.02 dex is almost negligible and the dispersion of 0.09 dex shows that our surface gravities results are precisely in line with the asteroseismic surface gravity data. 

\subsection{Comparison between GSN and LASP Parameters}

\begin{figure*}[tbp]
\centering
\includegraphics[width=\textwidth, angle=0]{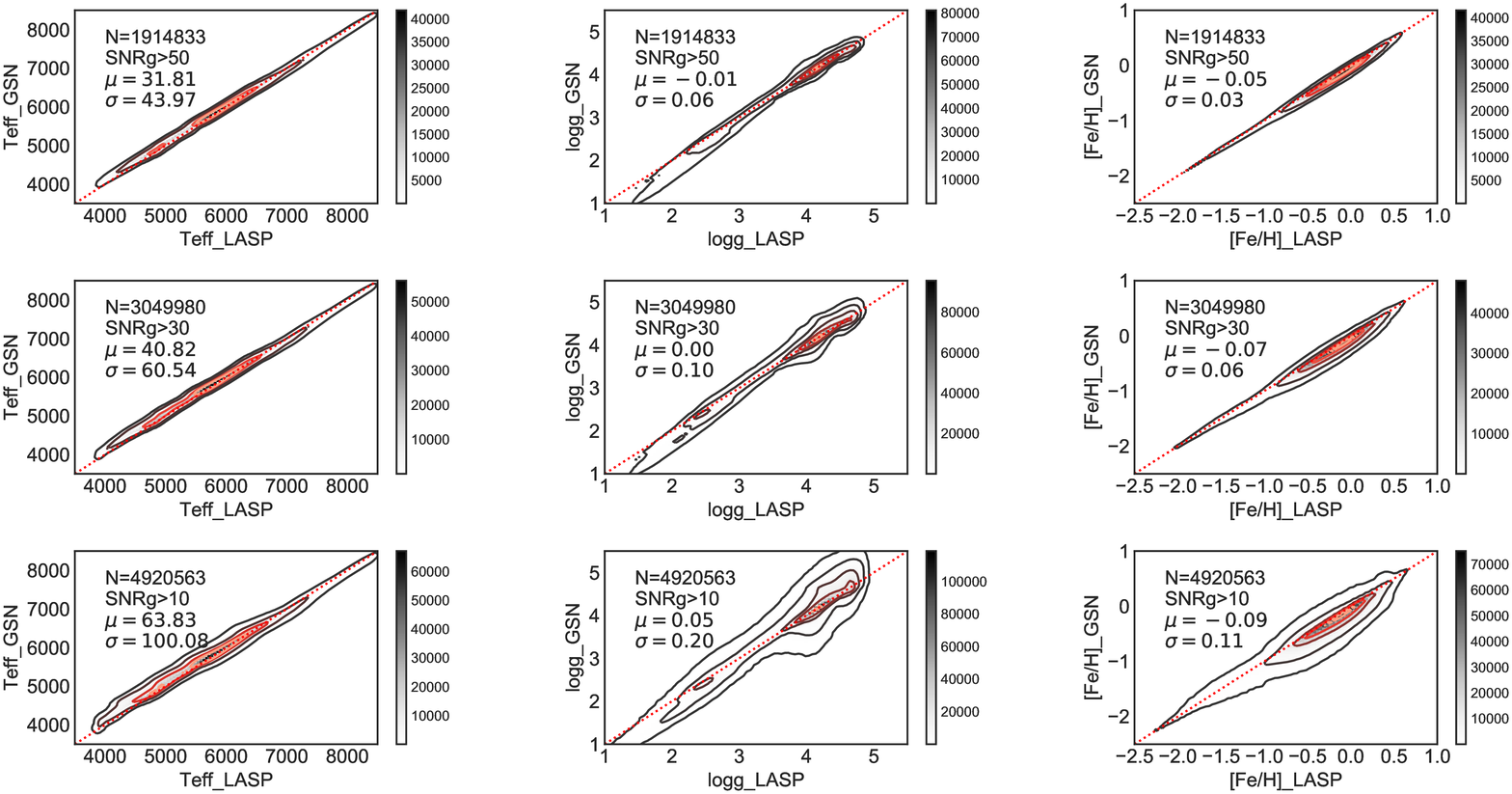}
\caption{The density distribution for comparison between GSN and LASP parameters ($\textit{T}_{eff}$ in left, log $g$ in middle and [Fe/H] in right panels) for LAMOST DR5 spectra with SNR$_{g}$ larger than 50, 30, 10 in the top, middle and bottom panels, respectively. The red dash lines are one-to-one line.}
\label{Figure7}
\end{figure*}

After comparing our results with the external stellar parameters, we finally chose the official parameter catalog of LAMOST, produced by the LAMOST Stellar Parameter Pipeline~\citep[LASP;][]{2015RAA....15.1095L,  2011RAA....11..924W} to compare and analyze the difference between them. LASP determines stellar parameters by employing two methods: the Correlation Function Initial (CFI) method is used to guess the initial values for the parameters, and the ULySS method~\citep{2011RAA....11..924W} is used to generate the final results. \citet{2011RAA....11..924W} highlighted that the precision values of LASP are 167 K, 0.34 dex, and 0.16 dex for $\textit{T}_{eff}$, log \textit{g} and [Fe/H], respectively, for early LAMOST observations. LASP required stars to be processed using the LAMOST 1D pipeline classified as late A, F, G, or K, the $g$-band SNR of SNR$_{g} \ge 15$ for bright nights and the SNR$_{g} \ge 6$ for dark nights~\citep{2015RAA....15.1095L}. We selected spectra with different SNR$_{g}$ levels to compare our results and the corresponding LASP parameters. As shown in Fig~\ref{Figure7}, good consistency is observed for the entire three stellar parameters at different SNR$_g$ levels, with little offset and small dispersion. Even at SNR$_g\ge10$, the dispersion is kept at 100 K for the effective temperature, 0.2 dex for gravity and 0.11 dex for [Fe/H]. Also, it can be determined that the deviation of the difference of the parameters decrease gradually as the SNR$_g$ increases. GSN seems to overestimate temperature while there is an underestimation of [Fe/H] with respect to the LASP parameters. 

\subsection{Comparison with other Spectral Fitting Methods}
\begin{figure*}[tbp]
\centering
\includegraphics[width=\textwidth, angle=0]{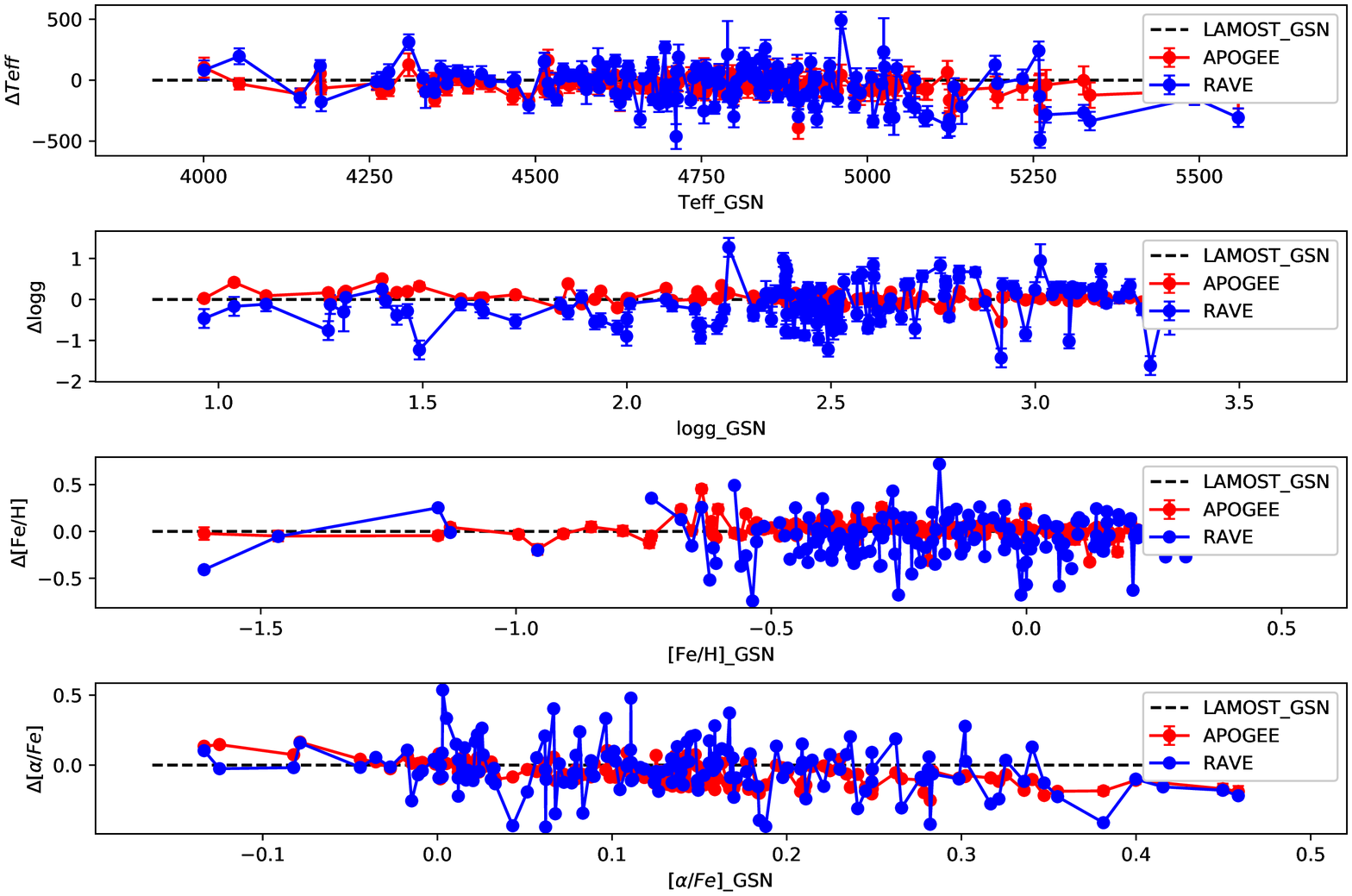}
\caption{The differences and errorbars of the stellar parameters of 339 common stars of LAMOST DR5, APOGEE DR14 and RAVE DR5 are shown, ($\textit{T}_{eff}$ is in the top panel, log $g$ and [Fe/H] are in middle two panels, and [$\alpha$/Fe] is in bottom-panel). The black dash lines in these panels are the value of zero which represent the GSN four parameters. The red and blue dots are the differences with the GNS parameters respectively. Please note that RAVE DR5 does not provide errors for [Fe/H] and [$\alpha$/Fe] so the scatters rather than errorbars are presented.}
\label{Figure8}
\end{figure*}

For comparing the results derived by GSN and independent spectral fitting methods, we employed a sample of common stars in 3 surveys: LAMOST DR5, RAVE DR5, and SDSS/APOGEE DR14. There are totally overlap 339 stars with stellar parameters determined by respective pipelines based on spectral fitting with empirical or theoretical spectra. The comparison results are showed in~\ref{Figure8}. We found that GSN's results show good agreements with both APOGEE's and RAVE's results. Moreover, our parameters are more consist with APOGEE's than RAVE. The reason is that the resolution (22,500) of APOGEE is higher than that (7500) of RAVE, while the wavelength range of LAMOST (3800\AA $\sim$9000\AA) is much wider than that of RAVE (8400\AA $\sim$8800\AA).

\subsection{Uncertainties}
\begin{figure}[tbp]
\centering
\includegraphics[width=0.5\textwidth, angle=0]{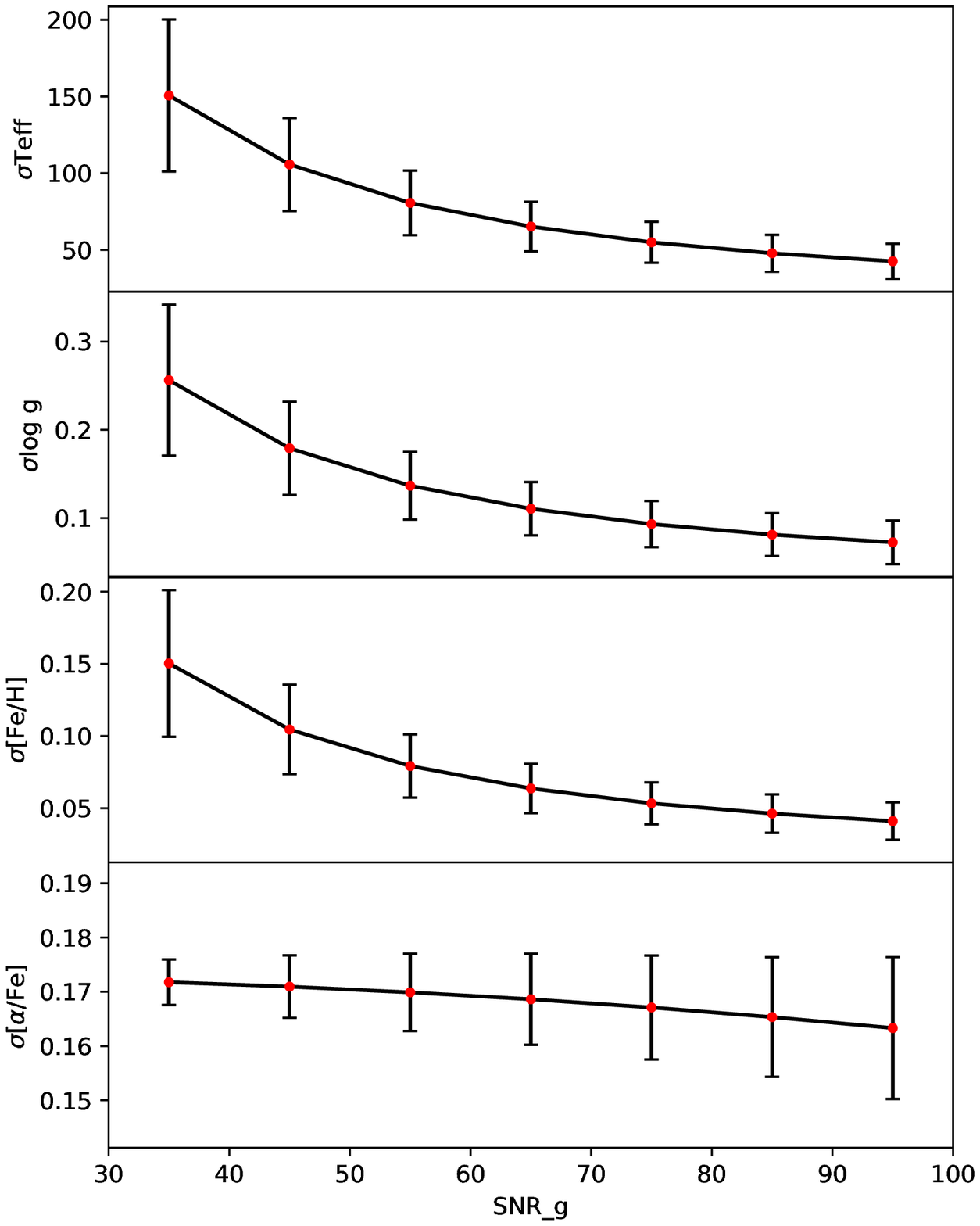}
\caption{The errors of the stellar parameters estimated by GSN for LAMOST DR5 spectra as a function of SNR$_{g}$ ($\sigma$Teff in the top panel, $\sigma$log g in the middle-top panel, $\sigma$[Fe/H] in the middle-top panel and [$\alpha$/Fe] in the bottom panel). The errorbar denotes the median value and the standard deviation of the errors in each SNR$_{g}$ bin. The SNR$_{g}$ bin borders are 30, 40, 50, ..., 100.}.
\label{Figure9}
\end{figure}

We generated estimates and error results for the parameters after the posterior distribution of the combination of the stellar atmospheric parameters ($\textit{T}_{eff}$, log $g$, [Fe/H] and [$\alpha$/Fe]) were obtained. Fig~\ref{Figure9} shows the mean and standard deviation of the errors associated with the stellar parameters as a function of SNR$_{g}$. It is evident that the errors have clear dependencies on the SNR$_{g}$ except for [$\alpha$/Fe]. These errors decrease rapidly with an increase in the SNR$_{g}$ when the value is less than 70, and a flat distribution beyond 70. The scatter of errors increases with SNR$_{g}$ for [$\alpha$/Fe] which is unusual. The mean uncertainties as shown in Fig~\ref{Figure9} are 150 K, 0.25 dex, 0,15 dex and 0.17 dex for $\textit{T}_{eff}$, log $g$, [Fe/H] and [$\alpha$/Fe] for the corresponding spectrum for SNR$_{g}$ larger than 30. The uncertainties decrease to 80 K, 0.14 dex, 0.07 dex and 0.16 dex when the SNR$_{g}$ of the spectrum is larger than 50. It was noted that the error is abnormal for [$\alpha$/Fe]. The reason for the irregular trend in the SNR$_{g}$ may be due to the fact that the initial values for [$\alpha$/Fe] cannot effectively constrain the real value for local 3-{$\sigma$} coverage, so that an incomplete posterior distributions for the abundance of metal elements with respect to hydrogen were obtained. Another reason is that utilizing all the flux from 3800$\AA$ to 9000$\AA$ increases the random error when fitting the observed spectrum to the model template. Most previous studies attempted to estimate [$\alpha$/Fe] by fitting a template to feature bands instead of the full spectrum~\citep{2011AJ....141...90L, 2016RAA....16..110L}. Therefore, the error for [$\alpha$/Fe] appears to be an serious overestimation which corresponds to the standard deviation of the external comparison.

\section{Discussion}
\label{sect:discussion}
A discussion on the error sources of the results is presented in this section. In general, the following aspects need to be carefully considered.

\begin{enumerate}

\item The pre-treatment. For pre-processing of the spectra, we simply normalize the training set and LAMOST spectra by dividing them by the median of their fluxes instead of dividing by the pseudo-continuum as outlined in SSPP~\citep{2008AJ....136.2022L}  and LASP~\citep{2015RAA....15.1095L}. We expect to exclude the error brought by pseudo-continuum normalization. The flux calibration and splicing process of the blue and red arms of the LAMOST spectra can increase the peak of the continuum. This accounts for the apparent overestimation of the temperature compared to the results derived based on the line feature (pseudo-continuum normalized spectrum). 

\item The offsets of parameters from different theoretical models are inherent, for example, NextGen (Phoenix) \textit{T}$_{eff}$ estimates are on average 2\% higher than ATLAS~\citep{2004AJ....128..829B} . The GSN training was based on the new version of the Phoenix model spectra,  which included some changes with respect to the old version including the treatment of the chemical equilibrium~\citep{2013A&A...553A...6H}.  Even if `perfect' models are assumed, differences would still exist between different sets of model spectra, which would lead to offsets comparing results based on different models. 

\item Sensitivity of  [$\alpha$/Fe] to spectral features in low resolution spectra. It is determined that [$\alpha$/Fe] are not as sensitive as effective temperature and surface gravity, and is a similar conclusion to that arrived upon in~\citet{2017ApJ...836...77Y}. This led to obvious offsets and dispersion with respect to results of APOGEE/ASPCAP and other catalogs. We will try to investigate the application of the maximum likelihood probability in the feature space in a future work.

\item The micro-turbulence speed and rotation of stars. We did not take the impact of micro-turbulence speed and the rotation of stars into consideration in this work. This is because the resolution of the LAMOST spectra is too low to constrain the micro-turbulence and rotation of stars, $v_{micro} ~and~ v_{rot}$, which may affect the accuracy of matching between the generated synthetic spectra and the observed spectra.

\end{enumerate}

\section{Summary }
\label{sect:summary}

We have designed a new structure of networks to generate labeled simulative LAMOST spectra and estimated the stellar atmospheric parameters (\textit{T}$_{eff}$, log \textit{g}, [Fe/H] and [$\alpha$/Fe]) of about 5.3 million spectra from LAMOST DR5 based on GSN combined with Bayesian theory by exploiting a distributed computing platform--Spark. Then, we utilized some sub-sample catalogs of LAMOST DR5 common stars with external precise stellar parameters derived from high-resolution observations, or by other methods, to ensure the reliability of our results. Our results show good consistency with the results from other survey and catalogs, although some small system errors were observed in a few instances. It was determined that our temperature scale was 100 K hotter than that of the normal spectroscopic temperature scales, similar to the scale of IRFM. Our surface gravity kept the height values consistent with the asteroseismic results. The system errors of [$\alpha$/Fe] were determined by an external comparison, in that these values are not sensitive to Bayesian probability fitting of the entire spectral bands. The precision of the parameters are listed as 80 K for \textit{T}$_{eff}$, 0.14 dex for log \textit{g}, 0.07 dex for [Fe/H] and 0.168 dex for [$\alpha$/Fe], for spectra with a SNR$_{g}\ge50$. The catalog is available on the internet at \url{http://paperdata.china-vo.org/GSN_parameters/GSN_parameters.csv}

\section*{Acknowledgement}
This work is supported by the National Key Basic Research Program of China (Grant No. 2014CB845700), the National Natural Science Foundation of China (Grant No. 11390371). The Guo Shou Jing Telescope (the Large Sky Area Multi-Object Fiber Spectroscopic Telescope, LAMOST) is a National Major Scientific Project built by the Chinese Academy of Sciences. Funding for the project has been provided by the National Development and Reform Commission. LAMOST is operated and managed by National Astronomical Observatories, Chinese Academy of Sciences.

\software{
		Numpy\citep{oliphant_guide_2006},
		Scipy\citep{jones2001scipy},
		Matplotlib\citep{Hunter:2007},
		Pandas\citep{pythonpandas},
		Keras\citep{chollet2015keras},
		Tensorflow\citep{tensorflow2015-whitepaper},
		Astropy\citep{astropy:2018},
		Spark\citep{Zaharia:2016:ASU:3013530.2934664}
		} 

\bibliography{paper_wangr}

\end{document}